\documentclass[10pt,a4paper]{article}
\usepackage{opex3}
\usepackage[utf8]{inputenc}
\usepackage{amsmath,amssymb}
\usepackage[amssymb]{SIunits}
\usepackage{graphicx}
\usepackage{bm}
\usepackage{cite}

\usepackage{color}

\newcommand{\suprm}[1]{\ensuremath{^{\text{#1}}}}
\newcommand{\subrm}[1]{\ensuremath{_{\text{#1}}}}
\newcommand{\supIn}{\suprm{in}}
\newcommand{\supOut}{\suprm{out}}

\newcommand{\Si}[1][i]{\vec S_{#1}\suprm{in}}
\newcommand{\So}[1][j]{\vec S_{#1}\suprm{out}}
\newcommand{\T}{\suprm T}
\newcommand{\SoT}[1][j]{(\So[#1])\T}

\newcommand{\alphai}[1][i]{\alpha_{#1}\suprm{in}}
\newcommand{\alphao}[1][o]{\alpha_{#1}\suprm{out}}
\newcommand{\rhoi}{\rho\suprm{in}}
\newcommand{\rhoo}{\rho\suprm{out}}

\renewcommand{\vec}[1]{\bm{#1}}
\newcommand{\vech}[1]{\vec{\hat{#1}}}

\renewcommand{\figurename}{Fig.}
\newcommand{\figref}[1]{\figurename~\ref{#1}}

\newcommand{\eqRef}[1]{Eq.~\eqref{#1}}
\newcommand{\EqRef}[1]{Eq.~\eqref{#1}}

\DeclareMathOperator{\Tr}{Tr}

\renewcommand{\Im}{\operatorname{Im}}
\renewcommand{\Re}{\operatorname{Re}}

\begin{document}

\title{The polarization properties of a tilted polarizer}

\author{Jan Korger$^\ast$, Tobias Kolb, Peter Banzer, Andrea Aiello, Christoffer Wittmann, Christoph Marquardt, and Gerd Leuchs}

\address{Max Planck Institute for the Science of Light, Guenther-Scharowsky-Str. 1/Bldg. 24, 91058 Erlangen, Germany\\
Institute for Optics, Information and Photonics, University Erlangen-Nuremberg, Staudtstr. 7/B2, 90158 Erlangen, Germany
}

\email{$^\ast$jan.korger@mpl.mpg.de}

\begin{abstract}
Polarizers are key components in optical science and technology. 
Thus, understanding the action of a polarizer beyond oversimplifying 
approximations is crucial. In this work, we study the interaction of 
a polarizing interface with an obliquely incident wave 
experimentally. To this end, a set of Mueller matrices is acquired 
employing a novel procedure robust against experimental 
imperfections. We connect our observation to a geometric model, 
useful to predict the effect of polarizers on complex light fields.
\end{abstract}


\section{Introduction}

Electromagnetic radiation is described as a vector field and, thus,
the orientation of the electric field vector, known as polarization, is of great importance,
both in classical and quantum optics. Polarized states of the light field
are often prepared and measured using polarizers and analyzers, respectively,
which can refer to the same device.
The physical implementation of such polarizing elements can be
very different
according to the application the device is designed for.
For example, the liquid crystal display (LCD) industry has refined their polarizer
design over the past decades to achieve the outstanding performance that these 
devices show today \cite{Hong2005,Hong2005b,Moon2010}.

From a more fundamental point of view, it is desirable to
work with a generic polarizer model, which is computationally convenient and
takes into account the geometric nature of the problem while being
suitable to describe a wide range of polarizers.
Such geometric models are currently used in the theoretical literature
\cite{Fainman1984,Aiello2009_FSpol}.
However, to the knowledge of the authors, 
they lack experimental validation, in particular for unusual corner cases.
Even for wide-view LCDs, the propagation angle inside the polarizing element
is not as steep as in our measurements.
Since any device which qualifies as a polarizer acts similarly on
a normally incident light beam, these obliquely incident waves
can be used to establish a realistic geometric model and demonstrate its validity.

In this article, the Mueller matrix of a commercial polarizer
made of elongated nano-particles shall be measured.
Reconstructing such a matrix from
potentially noisy experimental data is challenging and prone to
errors \cite{Aiello2006}.
We solve this problem using a self-calibrating polarimeter,
which additionally warrants that the result is physically acceptable
\cite{Anderson1994,Aiello2005}.
Our method combines a number of ideas discussed in the literature
\cite{Azzam1989,Goldstein1992,Schaefer2007,Branczyk2012}.

This article is structured as follows:
First, we introduce and illustrate polarization and polarizer models.
Then, we propose a Mueller
matrix polarimeter, which is robust against experimental imperfections
and does not rely on precision optics nor calibrated reference samples.
Finally, we employ this setup to obtain Mueller matrices for a commercial
polarizer and connect the observation to its microscopic structure. 
Motivated by recent theoretical and experimental work connecting the
action of a tilted polarizer to a beam shift phenomenon
\cite{Korger2011,Korger2013}, we extend our studies to include
the unusual case of almost grazing incidence.

\section{Polarization of a light beam}

In this work, we use both, Jones and Mueller-Stokes calculi, to
represent the polarization properties of the light field.
There are two key differences between both approaches.
First, the former method works with the electric field, while the latter
depends only on intensities, which can be directly measured.
And second, the Mueller-Stokes representation is more general since it
allows for describing
unpolarized states of light and depolarizing optical elements.

For our purpose, a collimated, polarized light beam can
be approximated as a planar wave field. A plane wave is completely determined by
the complex envelope $\vec J = E_x\vech x + E_y \vech y$
of its electric field $\vec E(\vec r,t) = \Re\left[\vec J\,\exp(i(\vec k\cdot\vec r - \omega t))\right]$,
where $\vec k = k\vech z$ is the wave vector.
The complex column-vector $\vec J$ has become known as the \emph{Jones} vector
\cite{Jones1941,Damask}.

Alternatively, the state of polarization of any light beam can be described by
a set of four real \emph{Stokes} parameters \cite{BornWolf}
\begin{align}
\begin{pmatrix}
S_0\\
S_1\\
S_2\\
S_3
\end{pmatrix}
=
\begin{pmatrix}
I_{0\degree} + I_{90\degree}\\
I_{0\degree} - I_{90\degree}\\
I_{+45\degree} - I_{-45\degree}\\
I_{R} - I_{L}
\end{pmatrix}
=
\begin{pmatrix}
|E_x|^2 + |E_y|^2\\
|E_x|^2 - |E_y|^2\\
E_x E_y^\ast + E_x^\ast E_y\\
i(E_x E_y^\ast - E_x^\ast E_y)
\end{pmatrix}
\text,
\end{align}
where $I_{\alpha}$ is the intensity of the light beam
transmitted across a linear polarizer
oriented at an angle $\alpha$ with respect to the $\vech x$
axis and $I_{R,L}$ is the right- or left-handed circularly polarized
component of the intensity.

The four Stokes parameters $S_\mu$ are related to the Jones vector $\vec J$
through the dyadic product of the Jones vector and its 
conjugate transpose
$\vec J^\dagger$ multiplied with the Pauli matrix $\sigma^{(\mu)}$
corresponding to each Stokes parameter \cite{Kim1987}:
\begin{equation}
    \vec S_\mu = \Tr\left[\left(
    \vec J \otimes \vec J^\dagger
    \right)
    \sigma^{(\mu)}
    \right]
    \text.
\end{equation}
The trace operation is in general irreversible.
Obviously, every Jones vector
$\vec J = (E_x,E_y)\T$ can be converted into a set of Stokes parameters,
whereas the reverse is not true.
In this article, we choose a basis
\begin{equation*}
    \sigma^{(0)} = \frac1{\sqrt2}\begin{pmatrix}
        1 & 0\\
        0 & 1
    \end{pmatrix},
    \sigma^{(1)} = \frac1{\sqrt2}\begin{pmatrix}
        1 & 0\\
        0 & -1
    \end{pmatrix},
    \sigma^{(2)} = \frac1{\sqrt2}\begin{pmatrix}
        0 & 1\\
        1 & 0
    \end{pmatrix},
    \sigma^{(3)} = \frac1{\sqrt2}\begin{pmatrix}
        0 & -i\\
        i & 0
    \end{pmatrix},
\end{equation*}
consistent with the definition of the Stokes parameters used in popular
textbooks \cite{BornWolf}.

Both approaches allow for a matrix calculus to describe linear operations
affecting the state of polarization \cite{Damask},
\begin{align}
    \vec J\supIn\rightarrow \vec J\supOut &= T\,\vec J\supIn\text,\\
    \vec S\supIn\rightarrow \vec S\supOut &= M\,\vec S\supIn\text,
\end{align}
where $T$ and $M$ are called \emph{Jones} and \emph{Mueller} matrices,
respectively. 

\section{Geometric Polarizer Models}

Generally, a polarizer is understood to project the light field onto a particular state of polarization.
For a plane wave impinging perpendicularly onto a linear polarizer, this state is trivially given by
the orientation of the polarizing axis.
In any other case, we need to work with a suitable model taking into account the physical nature of the
interaction.
For polarizers, for which the polarizing effect takes place at an interface
between two media,
e.g.\ reflection at the Brewster angle,
this problem is solved by applying the well-known boundary conditions or Fresnel formulas.

\begin{figure*}
	\begin{center}
		\includegraphics[width=\linewidth]{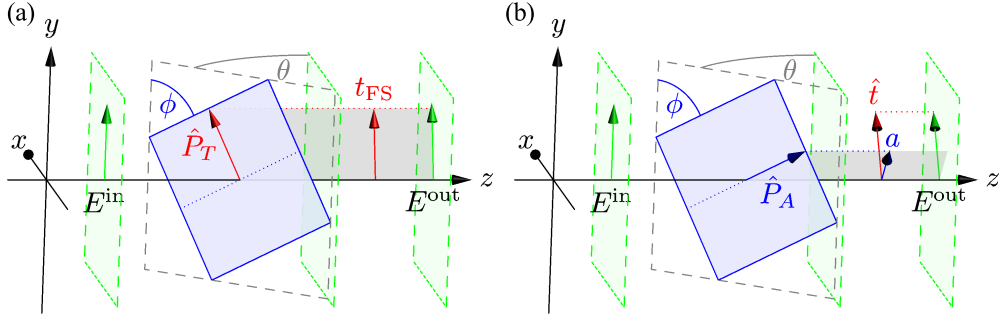}
	\end{center}
	\caption{
	Geometric interpretation of polarizer models.
	A plane wave with its electric field
	$\vec E\supIn$ in the $\vech x\vech y$-plane interacts with a 
	tilted polarizer not parallel to this plane.
	Our goal is to connect the orientation $\theta$, $\phi$ of the polarizer to the
	direction of the transmitted field component $\vec E\supOut$.
	(a) Fainman and Shamir \cite{Fainman1984} suggested to find this direction $\vec t\subrm{FS}$
	by projecting a vector $\vech P_T$ interpreted as the polarizer's
	\emph{transmitting axis} onto the $\vech x\vech y$-plane.
	(b) 
	The polarizer in question is made of elongated particles, all with their long
	axes oriented in direction of $\vech P_A$.
	Thus, our absorbing model makes of use of the projection $\vec a$ of the
	\emph{absorbing axis} $\vech P_A$.
	The field component parallel to $\vec a$ is scattered and eventually absorbed.
	Consequently, the transmitted field is polarized in
	direction of $\vech t$, orthogonal to $\vec a$.
	}
	\label{fig:geometry}
\end{figure*}

Fainman and Shamir (FS) have constructed a convenient geometrical model applicable to
polarizers that do not change the direction $\vech z$ of wave propagation
\cite{Fainman1984}.
They allow for an arbitrary orientation of the polarizer and assert that it can be completely
described with a three-dimensional unit vector $\vech P_T$.
FS make use of the transversality of the electric field vector
and
conclude that the effect of a polarizer reduces to the projection onto an
\emph{effective transmitting axis}
$\vech t\subrm{FS}$ (illustrated in \figref{fig:geometry}(a)).
In their model, the unit vector $\vech t\subrm{FS} \propto \vech P_T - (\vech z\cdot\vech P_T)\vech z$
is found by projecting the polarizer's transmitting axis $\vech P_T$ onto the
plane of the electric field perpendicular to the direction of wave propagation $\vech z$.

Fainman and Shamir's approach is practically useful since establishing
an effective transmitting axis
reduces the complexity of the intrinsically three-dimensional problem to
an operation on the two-dimensional Jones vector $\vec J$.
For any orientation of the polarizer, the resulting Jones matrix
$T\subrm{FS}=\vech t\subrm{FS}\vech t\subrm{FS}\T$ is a projector
as expected for an ideal polarizer.
However,
their recipe does not take into account the physical nature
of the interaction.

In this work, we attempt to 
adapt FS's approach to our observation.
In particular, we study a polarizing element made of 
anisotropic absorbing and scattering particles.
The ensemble of these elementary absorbers shall be oriented with their absorbing axes
$\vech P_A$ parallel to each other.
Analogously to the transmitting case, we interpret the projection of this unit vector $\vech P_A$
as an \emph{effective absorbing axis}
\begin{equation}
	\label{eq:vecha}
	\vech a = \frac{\vech P_A - (\vech P_A\cdot\vech z)\vech z}{\sqrt{1-(\vech P_A\cdot\vech z)^2}}\text.
\end{equation}
If this interpretation holds true, the light field after transmission
across an absorbing polarizer becomes
\begin{equation}
	\label{eq:E_absModel}
	\vec E\supIn \rightarrow \vec E\supOut = \vec E\supIn - (\vec E\supIn\cdot\vech a)\vech a\text.
\end{equation}
As above, the corresponding Jones matrix $T_A=1-\vech a\vech a\T = \vech t\vech t\T$ is
a projector, where $\vech t = \vech z\times\vech a$ can be interpreted as the
\emph{effective transmitting axis} as illustrated in \figref{fig:geometry}(b).
While \EqRef{eq:vecha} is structurally equivalent to Fainman and Shamir's construction,
our model coincides with their approach only for normal incidence.
Generally, our absorbing model $T_A=1-\vech a\vech a\T$ differs from the
FS case $T\subrm{FS}=\vech t\subrm{FS}\vech t\subrm{FS}\T$.
We want to note that the absorbing model can be found alternatively by treating the sub-wavelength
structure of the polarizer as a composite material, which behaves as an anisotropic
absorbing crystal \cite{Yeh1982}.

\begin{figure*}
	\begin{center}
		\includegraphics[width=\linewidth]{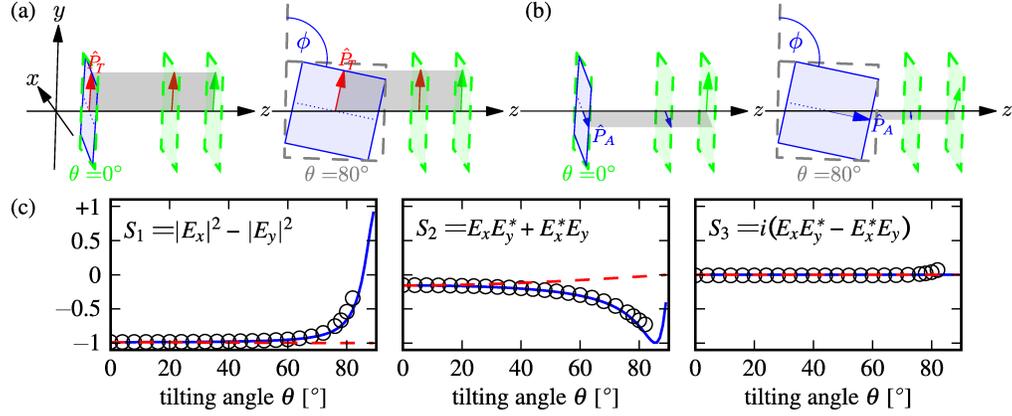}
	\end{center}
	\caption{
	State of polarization transmitted across
	a polarizer
	rotated around the vertical axis $\vech y$ by an angle $\theta$, keeping
	the angle $\phi=94.5\degree$ between the absorbing axis and
	$\vech y$ constant.
	(a) Visualization of the FS model \cite{Fainman1984}: The projection of
	the polarizer's transmitting axis $\vech P_T$ (red arrow) onto the plane of the
	electric field (green plane) determines the transmitted field component (green arrow).
	(b) Visualization of the absorbing polarizer model \eqref{eq:E_absModel}: The projection of
	the polarizer's absorbing axis $\vech P_A$ (blue arrow) onto the plane of the
	electric field (green plane) determines the absorbed field component.
	(c) Experimental data points (black circles) compared to both models.	
	The dashed red line depicts the original FS model, while the
	solid blue line describes the analogously constructed absorbing model.
	The data shows the polarizance vector $M_{i0}$ \cite{Lu1996} acquired as a part of
	our Mueller matrix measurement.
	This is the state of polarization after transmission across the polarizer
	if the incident wave is unpolarized.
	Only the absorbing model explains the
	drastic change of the transmitted state of polarization observed
	when the polarizer is tilted.
	}
	\label{fig:polAbsTrans}
\end{figure*}

We rely on empirical evidence to decide, whether
any of those two geometric models adequately describes our real-world polarizer.
To this end, we compare the state of polarization transmitted across the
polarizer to the one predicted by both models [\figref{fig:polAbsTrans}].
This shows that our polarizer can be approximated as a projector. When tilted,
the state of polarization, the device projects onto, is given by \eqRef{eq:E_absModel}.

This simple absorbing model, \eqRef{eq:E_absModel}, is the first main result of this article.
Using the Mueller matrix measurements reported in the following sections,
we can establish a phenomenological model and connect the observation
to a physical picture of the light field's interaction with the nano-particles.

\section{Mueller Matrix measurement}

\begin{figure}
	\centering
	\includegraphics[width=\linewidth]{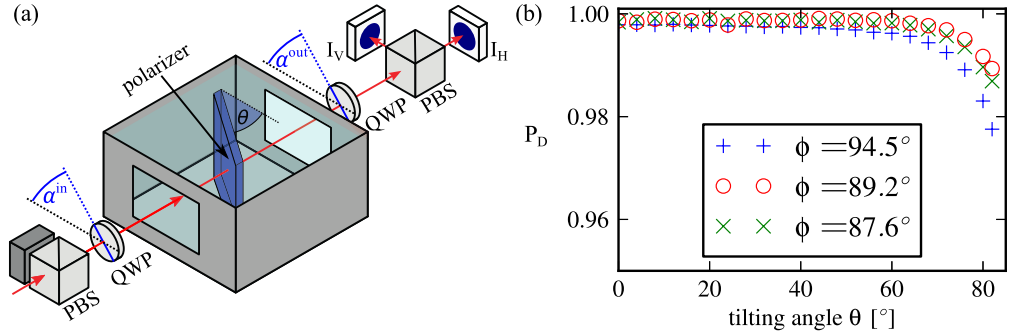}
	\caption{(a) Scheme of the Mueller matrix measurement.
	Using a collimated light beam (wavelength $\lambda = 795\,\nano\metre$),
	polarizing beam splitters (PBS), quarter wave plates (QWP), and two
	photo detectors I\subrm H and I\subrm V,
	the effect of an unknown sample on the polarization can be measured.
	For both QWPs, we use 6 different settings $\alpha\suprm{in/out}$ of their fast axes.
	Our sample is a commercial glass polarizer
	submerged in an index-matching liquid, which can be rotated around the
	vertical axis such that the incident beam impinges under an angle
	$\theta$.
	This setup allows to study the polarizing effect of the metal
	nano-particles, the polarizer is made of, without interference from the
	glass surfaces.
	(b) Observed depolarization index $P_D$ \cite{Gil1986} as a function of the orientation
	$\phi$, $\theta$ of the polarizer relative to the incident beam. $P_D = 1$
	describes a non-depolarization sample while $P_D = 0$ indicates a total depolarizer.
	}
	\label{fig:setup}
\end{figure}

In this section, we present a method to measure the  Mueller matrix of an arbitrary
optical element, which is robust against experimental imperfections,
such as noise and systematic errors.
With this least squares based estimation, we can gain full information about the
polarization properties of the device-under-test performing only a
limited number of intensity measurements.
The method employs a polarizer, a polarization beam splitter as an analyzer,
and two rotating wave plates to select the states of polarization
[\figref{fig:setup}(a)].

In any of these measurements,
the observed intensities
\begin{equation}
    \label{eq:Iij}
    I_{ij} = \frac12\,\SoT\,M\,\Si
\end{equation}
depend on the first waveplate,
which prepares a state of polarization $\Si$, the unknown Mueller matrix
$M$ describing the device-under-test, and the state of polarization $\So$,
we project onto at the detection stage.
Here, the row vector $\frac12\vec S^T$ describes the action of an analyzer
transmitting the state of polarization
given by $\vec S$. Applied to any Stokes vector, this yields the transmitted
intensity.

In principle, a generic real-valued $4\times4$ matrix $M$
is unambiguously determined by 16 equations like \EqRef{eq:Iij}.
However, the measured intensities $I_{ij}\suprm{E}$,
where the superscript $E$ denotes experimental values, can be noisy.
Thus, acquiring more than 16 values helps to reduce both statistical
and systematic errors significantly.
To this end, instead of solving a linear system of equations,
we pick the Mueller matrix $M\suprm{LS}$ from the set of
all possible Mueller matrices, such that
\begin{equation}
    \label{eq:sumSquares}
    \varepsilon(M\suprm{LS}) = \sum_{i,j}\left|
    \frac12\,\SoT\,M\suprm{LS}\,\Si - I_{ij}\suprm{E}\right
    |^2
\end{equation}
becomes minimal.

A Mueller matrix $M$ is physically acceptable \cite{Aiello2006,Anderson1994,Aiello2005} if and only if
its matrix elements
\begin{equation}
    M_{ab} = \Tr\left[H\left(\sigma^{(a)}\otimes{\sigma^{(b)}}^\ast\right)\right]
\end{equation}
are a function of a Hermitian matrix $H$ with non-negative eigenvalues \cite{Aiello2006}.
Any such matrix $H=H^\dagger$ can be expressed using a set of 16 real
numbers $\{h_1,\dots,h_{16}\}$.
Therefore, these 16 parameters span the vector space of physical Mueller
matrices and the set $\{h_1\suprm{LS},\dots,h_{16}\suprm{LS}\}$
which minimizes \EqRef{eq:sumSquares} yields to the best estimate
$M\suprm{LS}$ for the actual Mueller matrix.

If the states of polarization $\vec S$ are not known precisely,
we can find these parameter employing a procedure similar to the one described above.
Interestingly, this requires no
a priori information beyond the knowledge that the 
polarization states $\vec S$ are prepared using polarizers and birefringent retarders.
To this end, the device-under-test is removed from the beam path.
Using the same procedure as for the actual measurement, 
a set of intensities $I_{ij}\suprm{cal}$ is acquired, which characterizes the
setup.

Theoretically, this calibration run
corresponds to substituting $M\suprm{LS}$ in \EqRef{eq:sumSquares} with
the identity matrix (Mueller matrix of empty space).
Additionally, we express the abstract Stokes vectors
$\SoT = (\vec S_{H,V})^T\,M\supOut_j$
and $\Si = M\supIn_i\,\vec S_H$ in terms of Mueller matrices $M\supOut_j$ and
$M\supIn_i$, which physically describe our measurement device.
The Stokes vectors $\vec S_{H,V}$ represent horizontally or vertically
polarized states, respectively.
In our experiment, those are the states transmitted or reflected by a polarizing beam splitter
[\figref{fig:setup}(a)].
This yields:
\begin{equation}
    \label{eq:sumSquares_calib}
    \varepsilon(M\suprm{LS}) = \sum_{i,j}\left|\frac12(\vec S_{H,V})^T\,M_j\suprm{out}\,
    M_i\suprm{in}\,\vec S_H - I_{ij}\suprm{cal}\right|^2\text.
\end{equation}
In particular, $M\supIn_i(\alpha_i\supIn,\rho\supIn)$ represents the first
wave plate with the retardation $\rho\supIn$ and its fast axis oriented
at an angle $\alpha_i\supIn$ with respect to the $\vech x$ axis.
Analogously, $M\supOut_j(\alpha_j\supOut,\rho\supOut)$ describes the second wave plate.

Since we cannot fully rely on the manufacturer to specify retardation and
orientation of the fast axis with the desired accuracy, those values
are treated as unknown.
Nevertheless, employing motorized rotation stages, 
we can precisely reproduce relative movements $\Delta\alpha_i = \alpha_{i+1}-\alpha_i$
of both wave plates, where, for example, $\Delta\alpha_i=\Delta\alpha = 22.5\degree$.
Thus, our measurement setup is completely described by four parameters,
$\alphai[0]$, $\rhoi$, $\alphao[0]$, and $\rhoo$,
which are to be found with this calibration procedure.
The set of parameters
which minimizes \EqRef{eq:sumSquares_calib} yields
the states of
polarization $\Si$ and $\So$ relevant for our experiment.

As soon as these calibration parameters are known, 
\EqRef{eq:sumSquares} only depends on properties of the
device-under-test.
Minimizing \EqRef{eq:sumSquares} yields to the best experimental estimate
for the actual Mueller Matrix $M\suprm{LS}$ describing the device.

\section{Experiment}

In our experiment, we  study a polarizing interface,
made of anisotropically absorbing nano-particles.
To this end, we employ a commercial ``Corning Polarcor'' polarizer,
made of a glass substrates with 25 to 50\micro\metre\ thick polarizing layers
on each face.
These layers contain embedded, elongated and oriented silver particles.

We are particularly interested in the interaction beyond the trivial
case of normal incidence.
However, at larger angles of incidence $\theta$, the existence of
surfaces becomes problematic since a light beam propagating across
an interface experiences both, a change of its direction of propagation
(Snell's law) and of its polarization (a consequence of
Fresnel's formulas) \cite{BornWolf}.
These well-known effects are unrelated to the action of the actual
polarizing layer inside the glass substrate.
Thus, we avoid such surface effects by submerging the polarizer
in a tank filled with an index matching liquid (Cargille laser liquid
5610).
The refractive index of this liquid ($n_L=1.521$) matches with the one of
the polarizer's substrate
($n_G=1.517$).

Each measurement run consists of $6\times6$ steps, acquiring two
intensity values per step.
Every step uses a different combination of the
two wave plates' angles. %
For the required calibration run, we remove the polarizer from the beam path,
but keep the container with the index-matching liquid. 
Neither the liquid nor the glass windows were observed to affect the
state of polarization.
From this calibration data, we learn that both of our quarter-wave plates perform
within their specifications
($\alphai[0] = 2.57\degree$,
$\rhoi = \pi/2+0.008\,\radian$,
$\alphao[0] = 0.89\degree$, 
and $\rhoo = \pi/2+0.019\,\radian$).
Nevertheless, knowledge of these parameters is crucial to perform a
highly accurate Mueller matrix reconstruction.

Our goal is three-fold:
First, we attempt to establish a phenomenological model
taking into account the finite extinction ratio exhibited by real-world
polarizers.
Then, we demonstrate that this model accurately predicts the behaviour
for a wide range of parameters.
And finally, we connect the observation to the interaction of the light field
with the ensemble of nano-particles.

To this end, we perform three series of measurement runs for different orientations $\phi$
of the polarizer's absorbing axis, each for a large number of tilting angles
$0\degree\leq\theta\leq82\degree$.
We apply the least-squares method described above to find
the Mueller matrices describing our polarizer.

Results acquired with this method can be reproduced precisely.
Comparing independent measurements for the same configuration shows that
the statistical error of any Mueller matrix element $M_{ab}$ is less then
$10^{-3}$.
Furthermore, our data indicates that the results are also accurate.
The sample, we have studied is a linear polarizer.
For normal incidence ($\theta = 0\degree$), the transmittance across such a
polarizer does not depend on the helicity of the incident beam and the transmitted
beam is linearly polarized.
The corresponding Mueller matrix elements $|M_{03}|< 0.01$ and $|M_{30}|< 0.01$
clearly vanish for all relevant measurements.
Thus, we estimate systematic errors to be below $10^{-2}$.

\begin{figure*}
\includegraphics[width=\linewidth]{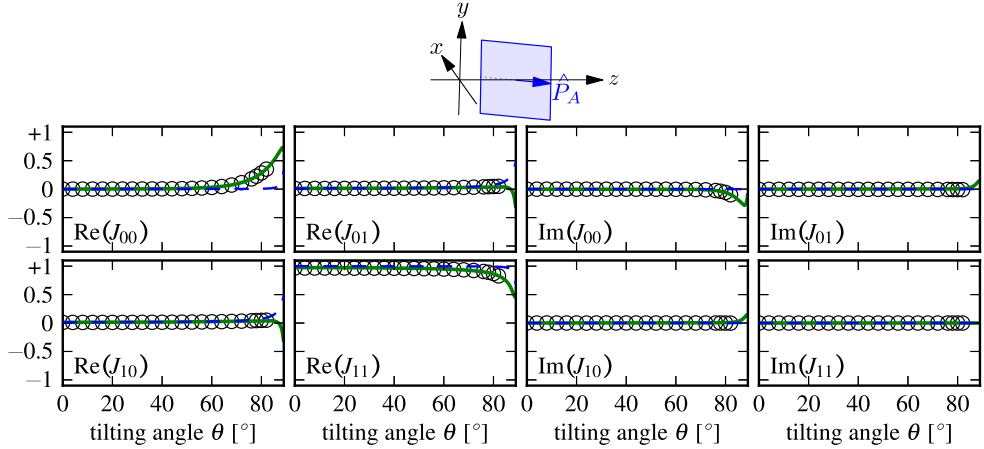}
 \caption{
    Jones matrix representation of the operation
    a light beam experiences when passing across our polarizer.
    The polarizer's absorbing axis $\vech P_A$ is oriented almost horizontally
    ($\phi = 89.2\degree$) and rotated around the vertical axis
    $\vech y$ by an angle $\theta$.
    The experimental data points (black circles) are calculated from our
    measured Mueller matrices.
    Ignoring an irrelevant global phase, we set $\Im(J_{11}) = 0$.
    Our phenomenological model, described by $T_P$, is depicted using solid green lines.
    Dashed blue lines show the geometric absorbing model given by $T_A$.
    }
 \label{fig:JonesData}
\end{figure*}

All measured Mueller matrices are practically non-depolarizing [\figref{fig:setup}(b)].
This means that 
a Jones matrix representation suffices to describe our sample.
The data series with the absorbing axis oriented almost horizontally
[\figref{fig:JonesData}] shows clearly that both the transmittance and the extinction ratios
decreases with the tilting angle $\theta$.
This behaviour cannot be described by a perfect projector as in the geometric
models discussed earlier.
Thus, we propose to generalize the projection rule, \EqRef{eq:E_absModel}, to include
two transmission coefficients $\tau_a$ and $\tau_t$ for states of
polarization parallel and perpendicular to the effective absorbing axis:
\begin{equation}
	\label{eq:E_pheno}
	\vec E\supIn \rightarrow \vec E\supOut = T_P\,\vec E\supIn\quad\text{ with }
	T_P = \tau_a\,\vech a\vech a\T + \tau_t\,\vech t\vech t\T\text.
\end{equation}

\begin{figure*}
\includegraphics[width=\linewidth]{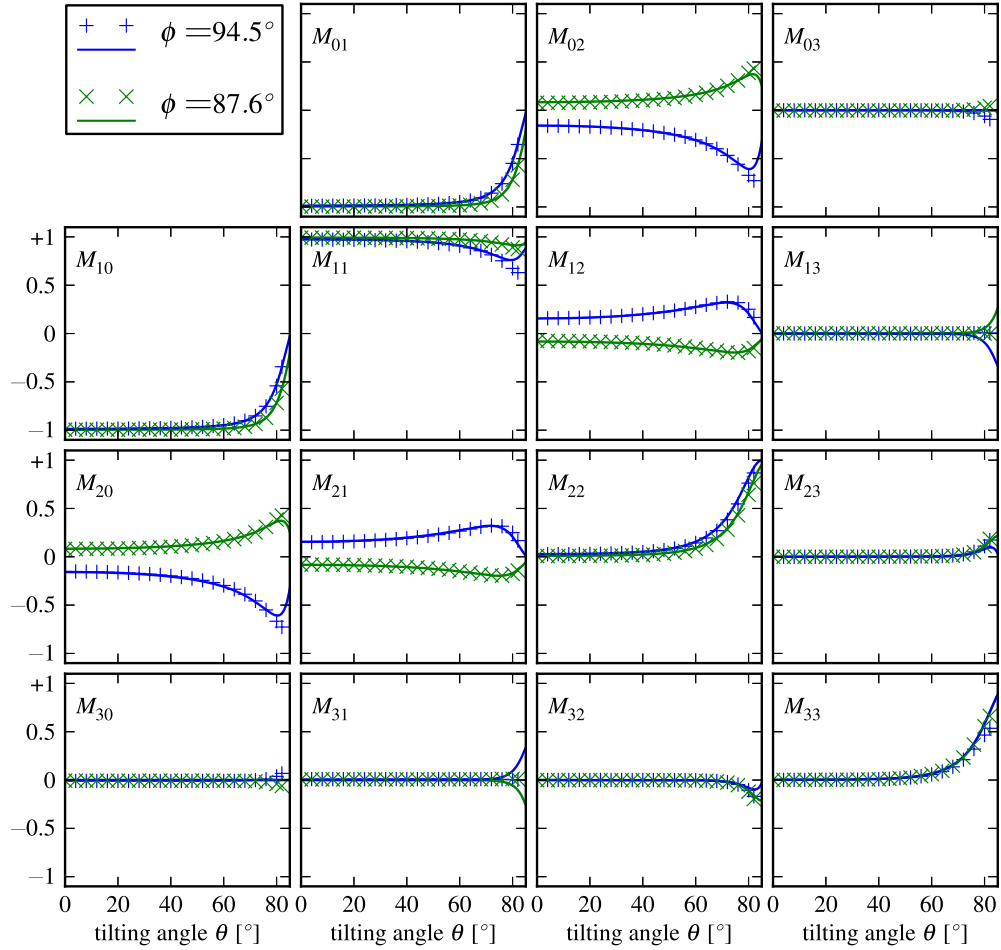}
 \caption{
    Reduced Mueller matrices $M' = \frac1{M_{00}}\,M$ describing
    the tilted polarizer for
    two different orientations of its absorbing axis
    $\phi$.
    Our polarizer model (solid lines) agrees well with the experimental data
    (markers).
    The model, we have employed, is deterministic.
    The small deviation from the model occurs for large
    tilting angles $\theta$, where the devices is slightly depolarizing
    (compare \figref{fig:setup}(b)).
    Depolarization effects cannot be modelled using Jones calculus as employed
    by our model.
    }
 \label{fig:MuellerData}
\end{figure*}

Using an ansatz implied by the qualitative behaviour of the data set shown in \figref{fig:JonesData},
we apply a curve-fitting algorithm to this data set, which yields:
\begin{subequations}
\label{eq:t}
\begin{align}
	\label{eq:tt}
	\tau_t(\theta) &= \exp(-0.025/\cos(\theta))
	\quad\text{ and}\\
	\label{eq:ta}
	\tau_a(\theta) &= 0.89\,\exp(-6.70\, \cos(\theta))
	-i\,0.62\,\exp(-13.6 \,\cos(\theta))\text.
\end{align}
\end{subequations}
Equations \eqref{eq:E_pheno} and \eqref{eq:t} constitute a phenomenological
model for our polarizer suitable to predict Jones and Mueller matrices
for any choice of the parameters $\theta$ and $\phi$.
In \figref{fig:MuellerData}, we demonstrate that this model
accurately agrees with our observation for different configurations.

Equation \eqRef{eq:tt} is a variant of
Beer's law \cite{BornWolf,Beer1852}
and describes how the absorption scales with the increasing effective thickness of the
sample when tilted.
The modulus square $|\tau_a|^2 > 0$ of \EqRef{eq:ta} accounts for
the transmittance for crossed polarization, i.e.\ the fact that even if the electric
field is polarized parallel to the effective absorbing axis, the absorption is
not 100\%.
The phase of the complex parameter $\tau_a$ indicates
that this field component is scattered with a phase determined by the
orientation of the nano-particles
relative to the incoming wave.

For small tilting angles $\theta<45\degree$, the observation
agrees with the prediction of the geometric absorbing model $T_A$.
Close to grazing incidence $\theta\rightarrow90\degree$,
the latter deviates, which we can understand in a physical picture.
The particles embedded in our polarizer are cigar-shaped \cite{Polizzi1997,Polizzi1998}.
Relevant for the polarization effect is the coupling of the light field to 
their long axes $\vech P_A$.
By design, the wavelength is close to the resonance of the particles' long axes.
At normal incidence, the scattering and absorption is strong
for states
of polarization parallel to the long axis
and negligible in the orthogonal case.

When the polarizer is tilted, only the component of the electric field vector
directed along the particles' absorbing axis $\vech P_A$ takes part in the interaction.
Thus, the effect of a single particle decreases proportionally to $\cos(\theta)$
as the coupling becomes less efficient.

The thickness of the polarizing layer guarantees that a light beam
interacts with multiple particles while propagating across the device.
Consequently, the observed extinction ratio is significantly larger than expected
for a single particle.
Our phenomenological model subsumes the sophisticated effect of this ensemble
using only two functions $\tau_a(\theta)$ and $\tau_t(\theta)$, which can be
directly measured.

\section{Conclusion}

We have presented a Mueller matrix polarimeter
making use of inexpensive linear polarizers and arbitrary retarding elements.
Our least squares optimization approach is fast, yet accurate
and precise. In particular, we have used this setup to study the effect a tilted polarizer has on the light
field.

Incidentally, linear polarizers are also popular as reference samples
to characterize such measurement devices.
Our data indicates that the combined statistical and systematic
error of any matrix element is less than $0.01$, while for polarimeters
of comparable speed and feasibility,
deviations between $0.03$ and $0.10$ per matrix element are typical \cite{Thompson1980}.
In fact, our method is comparable with the accuracy achieved by more sophisticated
calibration techniques requiring the use of multiple reference samples \cite{Compain1999}.

Finally, we have shown that a real-world polarizer, even when tilted, can be
modeled geometrically.
Using only the projection of the absorbing axis yielded already to
an acceptable approximation for the collective action of the nano-particle
ensemble.
It was demonstrated that the finite extinction ratio of realistic
polarizers can be taken into account phenomenologically,
including configurations close to grazing incidence.
We are confident that future work will connect the observation to a
detailed microscopic study of such nano-particles and their interaction with
the light field.

\section*{Acknowledgments}

The authors would like to thank Norbert Lindlein and Vanessa Chille for useful discussions,
and the anonymous Referees for insightful comments.

\end{document}